\documentclass[11pt,twoside]{article}
\usepackage{jltp}
\usepackage{graphicx}

\title{Influence of substrate charge on electron transport in narrow conducting channel}

\author{Yu.Z. Kovdrya, V.A. Nikolaenko, and A.V. Smorodin}
\address{B.I.Verkin Institute for Low Temperature Physics and Engineering , National
Academy of Sciences of Ukraine, 47 Lenin avenue, 61103 Kharkov,
Ukraine }

\runninghead{Yu.Z. Kovdrya et al.}{Influence of substrate charge
on electron transport in narrow conducting channel}
\begin{document}

\begin{abstract}
The electron transport in inhomogenous quasi-one-dimensional
conducting channels on the liquid helium surface are studied in
the temperature range $0.6-1.5  $ K. Inhomogeneities are created
by charging the substrate on which the conducting channels are
formed. It has been established that the electron conductivity
practically does not depend on temperature at some substrate
charge. The results obtained are explained by localization of
carriers and creation of electron polarons.
\end{abstract}
\maketitle

%Include this space if you do not use sections in your document.
%\vspace{0.3in}

\section{INTRODUCTION}
The investigations of electron transport in the system with
restricted geometry are of great interest\cite{1} . It was proved
to be possible to realize a quasi-one-dimensional (Q1D) charge
system with using surface electrons  on liquid helium (SE)\cite{2}
. It has been observed that character of the transport in
conducting Q1D channels essentially different in the channels of
different width and quality of the substrate surface on which the
conducting channels are formed. For a clean homogenous substrate
and narrow conducting channels, the electron mobility is
determined by interaction of carriers with helium atoms in vapor
and ripplons\cite{3,4}  and is described well by the
theory\cite{5}. As the width of the conducting channel increases,
an anomalous charge transport is observed\cite{6,7} ; the mobility
first increases with decreasing temperature \textit{T}, and then
at $T < 0.8$ K starts to decrease. The quality of the substrates
influences on the character of the electron transport. As it was
suggested, this effect is connected with appearance of potential
wells due inhomogeneities on the surface of the substrates. For
creation of potential wells it is convenient to use electron
charge deposited on the surface of a substrate. In this case it is
possible to change by a controlled way the density of the
potential wells. Preliminary data on the investigation of the
electron mobility in Q1D channels with charged surface have been
obtained in Ref. 7; in this work the results of detailed
investigations of the influence of the electron substrate charge
on the carrier conductance in Q1D channels over liquid helium are
presented.

\section{EXPERIMENTAL}

For the experimental investigation of conductivity, the cell
consisting of the top and bottom metal plates between which a
profiled dielectric substrate  is placed (Fig. 1).
\begin{figure}
%\centerline{\psfig{f1.eps,height=2.25in}}
\centerline{\includegraphics[width=0.70\textwidth]{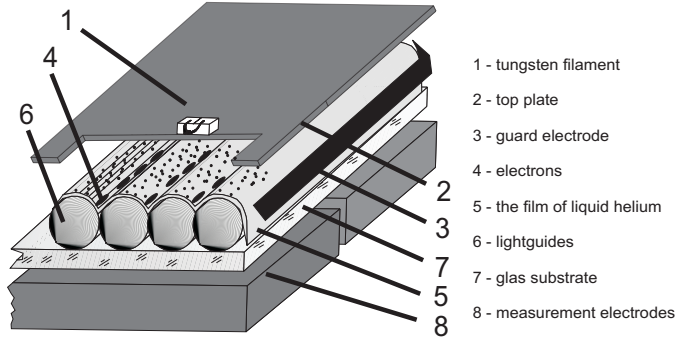}}
%\makebox[5in]{\rule[1.125in]{0in}{1.125in}}
\caption{Schematic of the experimental cell}
\label{fig1:yacheuka.eps}
\end{figure}
The bottom plate of the cell represents a coplanar system of
measuring electrodes consisting of two equal parts by size $ 6
\times 13.5 $ $ mm^{2} $ and with a gap between them \emph{a} $
\simeq 10 $ microns. The small value of the gap practically
excludes a nonuniformity of  electric field in a plane of the
substrate which is proportional to \textit{exp (-b/a)}, where
\textit{b} is distance from the lower plate to the electron
channels.

The structure of the substrate and technique of measurement of
conductivity were similar described in Ref. 4. A signal of
frequency 20 kHz from a generator was applied to the one of the
electrodes, the signal passing through the experimental cell was
taken off from the another electrode. The substrate represents a
50 lightguides of diameter 240 microns and length 12 \textit{mm}
located on a glass plate of the size $ 12 \times 13,5 \times 0,2 $
$mm^{3}$. Liquid helium covers the substrate realizing "deep"
liquid channels with some surface curvature  between the
lightguids. The tops of the lightguides are covered with the
liquid film of the thickness of about $300 \AA$. The shift of
amplitude of $ 0^{0} $ and $ 90^{0} $ components of an electrical
signal, passing through the cell with electrons, were measured;
this allowed to determine the real $ G_{r} $ and imaginary $ G_{i}
$ parts of the conductance of the cell.

Theoretical calculation of $ G_{r} $ and $ G_{i} $ with taking
into account the cell geometry are resulted in Ref. 4. For the
values of $ G_{r} $ and $ G_{i} $ the following expression has
been obtained:

\begin{equation}%\label{}
   \displaystyle G_{r} =n_{s} e^{2} \cdot \sum \limits
_{n,l=1}^{\infty }\frac{\Lambda _{n,l} e\omega ^{2} \chi _{1}
\lambda }{\left(m\omega _{q}^{2} -e\omega ^{2} \chi _{2} \lambda
\right)^{2} +(e\omega ^{2} \chi _{1} \lambda )^{2} }\,\,,
\end{equation}

\begin{equation}%\label{}
    G_{i} =n_{s} e^{2} \cdot \sum \limits _{n,l=1}^{\infty }\frac{\Lambda _{n,l}
\left(m\omega _{q}^{2} -e\omega \chi _{2} \lambda
\right)\omega}{\left(m\omega _{q}^{2} -e\omega \chi _{2} \lambda
\right)^{2} +(e\omega ^{2} \chi _{1} \lambda )^{2} }  +g_{0}\,\,.
\end{equation}
Here the value $ g_{0} $ characterizes conductance of the cell
without electrons, value $ \chi_{1} $ and $ \chi_{2} $ are real
and imaginary components of resistivity of the channels on one
electron, $ n_{s} $  is average density electrons in the cell, $
\omega_{q} $ is frequency of the plasma oscillations propagating
in system of parallel channels. The parameters $ \Lambda_{n,l} $
and $ \lambda $ in expressions (1) and (2) are determined by the
cell geometry. Estimations show that the main contribution to
value $ G_{r} $ and $ G_{i} $ is given by the first term.
Expressions (1), (2) allowed to determine conductivity $ \sigma$
of the channels and the value \textit{a}: $ \displaystyle
a=\frac{m\omega _{q}^{2} -e\omega \chi _{2} \lambda }{n_{s} }$.
The value  \textit{a} is determined by frequency of plasma
oscillations and the imaginary part of resistivity of the
channels. For free electrons
$m\omega^{2}_{q}>>e\omega\chi_{2}\lambda$, and the value
$a=\frac{m\omega _{q}^{2}}{n_{s}}$. Since $ \omega^{2}_{q}$ is
proportional to $n_{s}$ and inversely to \emph{m} (Ref. 4) the
value \emph{a} in this case depends only on the cell geometry.

\section{DISCUSSION}

\begin{figure}
%\centerline{\psfig{f1.eps,height=2.25in}}
\centerline{\includegraphics[width=0.52\textwidth]{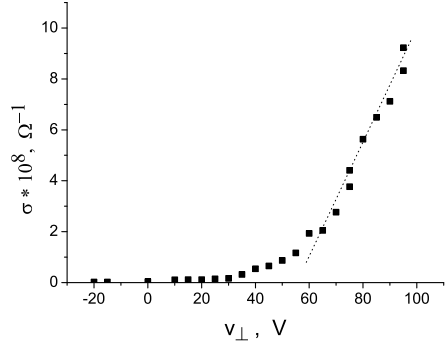}}
%\makebox[5in]{\rule[1.125in]{0in}{1.125in}}
\caption{A typical dependence of the conductivity of the channels
$\sigma$ on a pressing potential $ V_{\perp} $. }
\label{fig2:Graph1.eps}
\end{figure}

The results obtained are presented in Figs. 2, 3. A typical
dependence of the conductivity of the channels $\sigma$ on a
pressing potential $ V_{\perp} $ is shown in Fig.
\ref{fig2:Graph1.eps}.
The curve has been obtained in the condition of saturated charge.
It is seen, that the signal appears not with zero but from some
value $ V_{\perp cr} $. The value $ V_{\perp cr} $ since which the
signal appears is determined by density of electrons localized on
the helium film and on the substrate surface. In Fig. 3a
temperature dependence of the conductivity of the channels for the
electron density $ \sim 4\cdot 10^{12} $ $m^{-2}$ is presented. It
is seen, that $\sigma$ at first slightly increases with decreasing
temperature. At  \textit{T} $ \approx 0.9$  K a step changing of
the  value $\sigma$ approximately $10 \% $ is observed.
Unfortunately, the nature of such a decreasing $\sigma$ is not
clear at the  moment, and the additional experiments for the
explanation of this effect are needed. The week temperature
dependence of $\sigma$ allows to conclude that the helium atoms in
vapor and ripplons practically don't influence on the conductivity
of the channels.

In the region of temperatures 0.6 - 1.4 K the value \textit{a}
practically does not depend on temperature (Fig. 3b). As it
follows from Ref. 4, the theoretical value of \emph{a} depends
only on the cell geometry and must be independent on temperature.
The experimentally found values of \textit{a} is approximately two
orders more than the theoretical values of \textit{a} obtained for
the ideally conducting channels, and equals $ \sim 2.4\cdot
10^{-26} $ $ kg\cdot m^{2}/s^{2} $. The plasma frequency, which
has been calculated with using  the results of Ref. 4, is $ \sim
5.1\cdot 10^{8} $ $ s^{-1} $. The value $ \omega_{q}$, calculated
from the experimental value \textit{a}, is $ \omega_{q}$ = $
2.3\cdot 10^{9} $ $ s^{-1} $. It has been assumed in Refs. 3, 7,
that increase of the plasma frequency appears when localization of
particles in channels takes place. In this case, an additional,
so-called, "optical mode" appears; the value $ \omega_{q}$ thus
can be presented as   $\omega _{q} =\sqrt{\Omega _{0}^{2} +\omega
_{p}^{2} } $ , where  $\omega _{p} $ is plasma frequency of  ideal
conducting channels\cite{3,8}. Estimations with using of the
experimentally found value \textit{a }give for the $ \Omega_{0} $
the value $ \Omega_{0} = 2.3\cdot 10^{9} $ $ s^{-1} $.
\begin{figure}
%\centerline{\psfig{Graphic21.eps.eps,height=0.5in}}
\centerline{\includegraphics[width=0.9\textwidth]{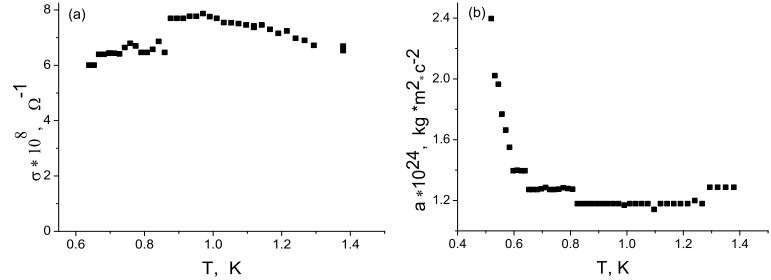}}
%\makebox[5in]{\rule[1.125in]{0in}{1.125in}}
\caption{The conductivity  of the channels  $\sigma$ (a) and the
value \textit{a} (b) as \emph{a} function of temperature.}
%\label{fig3:Graph2.eps}
\end{figure}
We can suppose, therefore, that electrons in quasi-one-dimensional
channels in presence of a charge on the substrate  are localized.
Characteristic value of the localization length of a particle is
$l_{loc}  \approx \left(\frac{\hbar }{m\Omega _{0} } \right)^{1/2}
=2.2\, \cdot 10^{-5} \, cm$. This value practically coincides with
average distance between electrons in the channels  $a_{0} \approx
n_{s}^{-1/2} =2.4\, \cdot 10^{-5} \, \, cm$.

One can supposed that such localization of particles can lead to
formation electron polarons: microscopic dimples on the liquid
helium surface with the effective mass noticeably exceeding the
mass of free electrons. The value $ n_{s} $ allows to estimate the
mobility of electron polarons in conducting channels, which is
equal $ \mu \approx 0.125 $  $ m^{2} / V\cdot s$. It is much less
than the theoretical value of the electron mobility in Q1D
channels which is $ \mu = 12 $  $ m^{2} / V\cdot s$ at $\textit{T}
= 1.4 $ K  \,\,\cite{5} . In the conducting Q1D channels above
liquid helium, with enough good homogeneity, a strong temperature
dependence of the electron mobility is observed \cite{4} . In this
connection, the week temperature dependence of $\sigma $, observed
in the Q1D channels at presence of a charge on the substrate, is
interesting. One of the possible explanations of this effect
consists of that due to the large number of the potential wells
formed by the immobile electrons, the carrier transport is
realized by tunneling of particles from one potential well to
another. The diffusion of particles in this case is determined by
the expression   $D\approx \nu _{0} \cdot a_{0} ^{2} \cdot \exp
[\int \limits _{a}^{b}\sqrt{2m\Delta V} dx ]$, where $\nu _{0}
=\frac{\Omega_{0} }{2\pi } $ is frequency of oscillations of an
electron in a potential well, $ \Delta V $ is height of the
potential well. The value $\textit{a}_{0}$  set the characteristic
scale of variations of the potential relief, it is determined by
average distance between the electrons localized on a thin helium
film and on the dielectric substrate. The appreciations, however,
show that, for an explanation of observable values of the
mobility, it is required to assume that $ \Delta V = 0.2 $ K. It
is too small value for realization of the tunnelling mechanism.
More probable explanation of observed features of the carrier
transport, is scattering electron polarons by variations of the
potential, created both localized immobile electrons, and
inhomogeneities of channels.

The electron transport in quasi-one-dimensional channels on the
liquid helium surface at presence of inhomogeneities on a
substrate was considered in Ref. 9. It has been shown that the
mobility of carriers limited by scattering of electrons by
defects, can be presented as
\begin{equation}%\label{}
\mu _{d} \approx \frac{e}{m\nu _{d} } \left(\frac{kT}{\hbar \omega
_{0} } \right)^{1/2}.
\end{equation}
Here  $ \nu_{d} $ is frequency of collisions electrons with
surface defects of the substrate, \textit{k} is Boltzman constant,
$ \hbar $ is Plank's constant, $ \omega_{0} $ is frequency of
oscillations of an electrons due to the movement of an electron
across the channel. As defects the roughness of the substrate on
which the conductivity  channels are formed were considered. It
was also supposed that a layer of liquid helium in a channel thin
enough. Besides, at calculation the condition $\textit{kT}<<\hbar
\omega_{0} $ was used. It has been established that the value
$\nu_{d} $ does not depend on temperature and consequently
temperature dependence of $\mu$ is enough weak: $ \mu \sim
T^{1/2}$. In conditions of our experiment $ \hbar\omega_{0}=0.2$
K, so the requirement $\textit{kT}<<\hbar \omega_{0} $ does not
take place. Beside this scattering of the electrons on variations
of the random potential, probably, has Coulomb character.
Nevethelees, apparently, one can expect that weak temperature
dependence of the mobility in this case should takes place.
Unfortunately, now there is no theory of the electron transport
under the scattering electrons by the random potential caused by
the charge of the substrate.

Thus, in the present work the electron transport in the
quasi-one-dimensional channels is investigated at presence of a
charge on the substrate on which the channels are formed. It is
shown that the conductivity of the channels practically does not
depend on temperature. It is made the assumption that localization
of carriers in the non-uniform channels, apparently, is
accompanied with formation of electron polarons. The effective
mass of such a polaron essentially exceeds the free electron mass.

\end{document}